\documentclass[fleqn,10pt]{wlscirep}
\usepackage{subcaption}
\usepackage{graphicx}
\usepackage[utf8]{inputenc}
\usepackage[T1]{fontenc}
\usepackage{textcomp}
\DeclareUnicodeCharacter{E5F8}{\ensuremath{\ominus}}
\title{A Modelling study of hole transport in GaN/AlGaN superlattices}

\author[1,*]{Mengxun Bai}
\author[1]{Judy Rorison}

\affil{Department of Electrical and Electronic Engineering, University of Bristol, Bristol, BS8 1UB, UK}

\affil[*]{mb18200@bristol.ac.uk}

\begin{abstract}
The transport of holes through p-doped wurtzite bulk GaN and AlGaN is poor so transport of holes through GaN/AlGaN superlattices has been proposed and investigated theoretically and experimentally with experimental results showing poor transport. The reason for this poor performance is not fully understood. In this paper, the transport of holes in GaN/AlGaN wurtzite crystal superlattices is investigated through theoretical modeling, examining the role of the composition of the Al$_x$Ga$_{1-x}$N barrier regions and the thickness of the GaN quantum wells and the AlGaN barriers in determining the position and width of the heavy hole miniband. To consider the transport of the holes in the miniband we examine the effective mass of the miniband and possible scattering mechanisms. In particular, ionized impurity(II) scattering from ionized acceptors in the barrier regions is investigated as it is deemed to be the dominating scattering mechanism degrading hole transport. The energy position of the miniband relative to the ionized impurities and the wavefunction overlap with the ionized acceptors in the barrier regions is investigated to minimize II scattering. Some designs to optimize hole transport through wurtzite p-doped GaN/AlGaN superlattices to minimize II scattering are proposed.
\end{abstract}
\begin{document}

\flushbottom
\maketitle
%
%
\thispagestyle{empty}

\section*{Introduction}

The transport of electrons and holes in GaAs/AlGaAs superlattices has been extensively studied using various device structures, such as resonant tunneling diodes, superlattice infrared photodetectors, and quantum cascade lasers[1][2][3]. These devices utilize the minibands to control the transport of electrons and holes and exhibit unique electronic properties. Similarly, it should be possible to grow GaN/AlGaN superlattices to have unique electronic and optical properties that can be tuned by adjusting the thickness and composition of the individual layers similar to the GaAs/AlGaAs system. In contrast to GaAs and AlGaAs, GaN and AlGaN have a wurtzite crystal structure resulting in different band structure, and also have other physical properties such as a wide bandgap making these materials particularly suitable for high power and high frequency applications[4][5][6]. In common with most wide-bandgap semiconductors the acceptor binding energy is very large (>100meV) making the activation of p-doping difficult  resulting in high p-resistivity. The idea of using superlattices to free the holes and exploit transport in a miniband in the perpendicular direction to the superlattice period could aid devices which require electrons and holes in an active region such as LEDs or lasers or electronic devices which  require hole transport such as PMOS[7-12]. This concept was patented by one of the authors for use in an LED/laser design[7] with a priority date of 1996. However grown and fabricated superlattices in GaN/AlGaN were found not to exhibit good perpendicular hole transport[7][8]. The aim of this study is to investigate why this is so. GaN/AlGaN superlattices have high internal electric fields arising from their wurzite band structure (piezoelectric fields) and spontaneous polarization from their interfaces which is different from the zinc blende GaAs/AlGaAs system and also have the very deep ionized acceptor levels. In this study we investigate how the miniband can be tuned by varying the barrier composition (low Al barrier content reduces the piezoelectric field) and the well and barrier thicknesses and investigate the miniband position and Fermi level relative to the acceptor levels.

This work was motivated by an investigation into GaN/AlGaN superlattices by Duboz (2014)[9] who examined these effects deciding that vertical hole transport will not be good through the superlattices. He restricted his investigation to equally sized Quantum Well (QW) and Quantum Barrier (QB) so we have examined varying QW and QB thicknesses($L_{QW}$ and $L_{QB}$) independently with the aim of creating a superlattice with a wide energy band and a large hole concentration in the superlattice. We then re-examined the role of II scattering in the superlattice with the aim of examining how different designs impact upon this effect. The tantilising promise of reduced resistivity and improved vertical hole transport in (Al)GaN/AlGaN superlattices is worth further study. We agree with Duboz that $L_{QW}$ and $L_{QB}$ larger than 8 monolayers(MLs) for each, corresponding to roughly 4nm thickness will result in multi-quantum well behaviour rather than superlattice behaviour which will not benefit vertical transport. Also $L_{QW}$ and $L_{QB}$ less than 4 monolayers(MLs) may result in an alloy rather than a superlattice so the focus on this paper will be on dimensions between these limits. Other considerations involve maintaining a continous miniband through the structure for the applied field rather than breaking up into Wannier-Stark ladders in which transport would be expected to be very poor and also not to have the condition of transport with Bloch oscillations which can arise in high fields with low scattering. Recent interest on employing superlattices based on both QW and QB comprised of AlGaN for application for application in UV emitters is currently being considered[10] and would be expected to be even more difficult to design. It is therefore important to continue to examine transport in p-doped GaN/AlGaN particularly as the material parameters are being reexamined and across multiple cases, offering a comprehensive understanding of these complex interactions.[11][13].

In this paper, the first contribution is the calculation of the band edges of Mg-doped GaN/Al$_x$Ga$_{1-x}$N superlattices with different aluminum compositions and QW and QB thicknesses, not restricting the QW and QB widths to be the same. The calculation considers the band shift due to strain, piezoelectric, and spontaneous polarisation effects and also includes space charge effects. We show how the energy position and width of the miniband can be designed by varying these parameters and how important is the Al concentration in the barriers in controlling the internal electric field in the superlattice and the miniband width. We examine the miniband dispersion where the effective mass of the miniband in the vertical direction can be used to determine low-field mobility. It is important that this does not get large reducing mobility. Then the concentration of free holes in the miniband can be examined and optimized as a function of $L_{QW}$ and $L_{QB}$ whilst maintaining a good energy width of the miniband. This value can be directly related to the resistivity and also impacts directly conduction laterally along the superlattice (similarly to HEMTs). Then vertical transport through the superlattice is considered where the effective mass of the hole in the superlattice direction, the wavefunction of the superlattice, and the scattering processes are considered. The two primary scattering mechanisms are hole-longitudinal optical (h-LO) phonon scattering and ioinzed impurity (II) scattering. In h-LO phonon scattering, holes transfer energy to phonons through a Coulomb interaction which results in a small deflection angle (or momentum change)[14-16]. Conversely, II scattering involves elastic scattering that changes the momentum direction of electrons without energy loss[17]. Therefore, as long as the miniband energy width is larger than the energy of the LO phonon scattering can occur within the miniband and not hinder the direction of travel significantly. If the miniband has an energy width less than the LO phonon this scattering process is suppressed as the initial and final state must be within the miniband (for intraband scattering). As the LO phonon energy is >50meV this is the case in these superlattices. Therefore we believe that h-LOphonon scattering should not be significantly different in GaN/AlGaN superlattices compared to GaAs/AlGaAs and anticipate that it should not be very detrimental. As a result, II scattering is often considered the main scattering mechanism and its effects on hole transport are critical to understanding and optimizing the hole transport[9][18][19]. The acceptors are deep, have a higher concentration than in GaAs/AlGaAs superlattices and have a non-uniform distribution-being ionized only in the barrier regions (even if not intentionally modulation doped). In this paper, we examine the role of II scattering and examine how we can control the miniband design to minimize it considering both a single particle scattering approach and considering the wavefunction of the superlattice and its probability of being in the barrier where the ionized acceptors are located (equivalent to the time taken for the holes to transit through the barriers). Based on our findings, we propose a set of GaN/Al$_x$Ga$_{1-x}$N superlattices designs that exhibit relatively low levels of II scattering.

\section*{Method applied on GaN/AlGaN superlattices}
Nextnano software is used to solve the Schrodinger wave equation for the periodic potential created by the alternating material layers of GaN and AlGaN grown along a growth direction z. It assumes an effective mass continuous medium approach rather than an atomistic localized energy level approach. The sharp energy levels of the individual QWs coalesce through the decaying of the wavefunction in the narrow QB material and broaden to form a miniband through the superlattice. Eigenfunctions are composed of the build-up of a series of plane waves $k_i$ which are continuous at the interfaces $\psi$ and smooth at the interfaces(conserving current) $\frac{1}{m^*}\frac{d\Psi}{dz}$[20]. The wavefunction must be periodic with the lattice period L: $\Psi(z+L)=\Psi(z)exp(ikL)$. The boundary conditions are periodic meaning the solution is for an infinite superlattice which is appropriate for more than 10 periods. The period L of the superlattice is the sum of the quantum well(QW) width ($L_{QW}$) and the barrier(QB) width ($L_{QB}$), denoted as L = $L_{QW}$ + $L_{QB}$. This superlattice calculation allows for calculating the miniband properties and corresponding energy levels.

In addition to the general superlattice band calculations which depend on energy band differences, effective masses of the highest level valence band states or lowest level conduction band states in the QW and QB materials and strain effects from lattice mismatch, the wurtzite also have in-built electric fields. Ionized dopants and free carriers also contribute to energy shifts and band bending. These effects are included in the NextNano software by coupling the Schrodinger wave equation to a Poisson solver and solving iteratively. Nextnano is employed to model the superlattice band edge and miniband including the non-linear Poisson equation solver which uses the iterative method of preconditioned conjugate gradient (PCG) for calculation[21]. The Poisson-Schrodinger equation is solved self-consistently with periodic boundary conditions, which involves an iterative solution procedure. The electrostatic potential and wave functions are updated iteratively until they achieve self-consistency, enabling the accurate calculation of the band edges and other properties. The solutions depend critically on the input parameters for the input materials for GaN and AlGaN.

Gallium nitride (GaN) and Aluminium gallium nitride (AlGaN) are polar when naturally grown in the wurzite crystal structure[9][22]. A table containing the physical constants used in this study for GaN and AlN are given in Appendix A.  All of the material parameters for Al$_x$Ga$_{1-x}$N except the energy gap are taken from the compositional weighting of the binaries. The energy gap is found from a quadratic expression given as shown in Appendix A. The wurtzite band structure results in an internal electric field, which can affect the electronic properties of heterostructures grown along the c-axis of the wurtzite crystal structure. In these heterostructures, the internal electric field leads to a potential triangular profile in the quantum wells and barriers. Therefore, it is crucial to consider the internal electric fields in the z direction: $F_z^W$ and $F_z^B$ in the QW and QB, respectively, when designing and modeling these heterostructures. The internal electric fields in GaN and related alloys are mainly caused by spontaneous polarization (SP) and piezoelectric polarization (PZ). SP arises from the asymmetric distribution of charges in the crystal structure, while PZ arises from the strain-induced polarization due to lattice mismatch between the layers in the heterostructure. To calculate the electric fields in the well and barrier due to SP and PZ polarization along the growth direction in p-type GaN/Al$_x$Ga$_{1-x}$N superlattices at different Al compositions, the following formulas can be used.

 \begin{subequations}

\begin{align}
F_z^W=&\frac{P_{sp}^b+P_{pz}^b-P_{sp}^w-P_{pz}^w}{\epsilon^w_r+\epsilon^b_rL_w/L_b} \label{5a}\\
F_z^B=&-\frac{L_w}{L_b}F_z^w\label{5b}
\end{align}
\end{subequations}

where $P_{sp}$ and $P_{pz}$ illustrate the normal polarization with respect to the growth plane of superlattices(0001). Their superscripts w and b correspond to the well and barrier regions in a superlattice structure. Additionally, $L_{QW}$, and $L_{QB}$ are the well and barrier thicknesses respectively while $\epsilon_r^w$ and $\epsilon_r^b$ are the relative static dielectric constants of the QW and QB, respectively. $P_{sp}$ and $P_{pz}$ are given as the functions of crystal orientation and Al composition. Park's study[23] listed the parameters and formulas for $P_{sp}$ and $P_{pz}$ polarization of GaN and AlN, and the results for Al$_{0.2}$Ga$_{0.8}$N were calculated. Based on this, we were able to calculate the polarizations for ternary alloys containing other aluminum compositions in the (0001) growth direction of different superlattice structures. The parameters used for these calculations were obtained from references [23] and [24]. The constants for the ternary alloys were determined through linear interpolation of the parameters for the corresponding binary alloys based on their compositions.

In addition to the internal electric fields discussed above there can also be electric fields caused by crystal distortions due to growth on lattice mismatched substrates. We consider GaN/AlGaN superlattices with the AlGaN barrier under tensile strain and the GaN well lattice-matched to the substrate (sapphire). According to reference [23][24], the tensile strain of the barrier under different Al compositions is calculated in Table 1.

\begin{table}[ht]
\centering
\begin{tabular}{|l|l|l|l|l|}
\hline
Al$_{0.2}$Ga$_{0.8}$N & Al$_{0.4}$Ga$_{0.6}$N & Al$_{0.6}$Ga$_{0.4}$N & Al$_{0.8}$Ga$_{0.2}$N & AlN \\
\hline
0.484\% & 0.97\% & 1.46\% & 1.963\% & 2.48\% \\
\hline
\end{tabular}
\caption{\label{tab:example}The tensile strain of the QB under different Al compositions}
\end{table}

In studying p-doped GaN/Al$_x$Ga$_{1-x}$N superlattices, determining the acceptor energy level is crucial for materials with different compositions. By incorporating the Mg acceptor energy level in the model, we can determine if the acceptors are ionized and their energy separation from the miniband. For GaN, the large Mg-acceptor ionization energy is around 170 meV. For Al$_x$Ga$_{1-x}$N, the acceptor energy level ranges from 170meV$\sim$517meV depending on composition. The activation energy $E_A$ of the Mg acceptor in Mg-doped Al$_x$Ga$_{1-x}$N as a function of the Al content x is given in the following formula[25].

 \begin{equation}
E_A = \frac{m_h^*}{m_0}\frac{E_0}{\epsilon_r^2}\label{4}\\
\end{equation}

Where $m_h^*$ and $m_0$ are the hole effective mass and electron rest mass, respectively. $E_0$ is a constant[23] and $\epsilon_r$ is the relative static dieletric constant[23]. 

The effective mass of the hole is an important parameter but relatively difficult to define. The effective mass of holes in GaN and AlN is very anisotropic and depend on the crystal direction. In the study, we have adopted parameters that are widely recognized in the literature[26,27] giving $m_x$=$m_x$=1.6$m_0$ (10.42$m_0$) while $m_v$=1.1$m_0$ (3.53$m_0$) for GaN (AlN). Since the superlattice structures involved in this paper are grown in the z-direction (0001), the hole mass in the z-direction is used in calculating the vertical superlattice miniband. Other parameters are taken from references [21-29] which are listed in Appendix A.

\section*{Results}
\subsection*{Miniband model}

In this study, we do a comprehensive parameter sweep of barrier composition and QW and QB widths not restricting to equal QW and QB widths, $L_{QW}$ and $L_{QB}$ respectively, while the period is taken to be L= $L_{QW}$+$L_{QB}$. In this study, we do not restrict that $L_{QW}$=$L_{QB}$. There are limits to material growth technology but bearing this in mind we can explore a wide parameter space to design minibands with features we want to optimize. Using the lattice constants of GaN and AlN 0.26nm and 0.25nm respectively, we design and present our superlattices in terms of monolayer thicknesses[28]. The lattice constant of the Al$_x$Ga$_{1-x}$N alloys are as taken to vary in a linear function between the two binaries. The QW and QB of interest in our study of GaN/Al$_x$Ga$_{1-x}$N range from 2MLs to 8MLs so are Short Period Superlattices (SPS).

The heavy hole (HH) and light hole (LH) minibands in two superlattice structures grown on a sapphire substrate are overlaid on the superlattice potential profiles in Fig.1 for two different superlattice structures with Mg p-doping of $10^{20}cm^{-3}$.  It is seen that the valence band experiences a band offset due to composition that the band edges experience significant band bending due to the effects of the piezoelectric field and spontaneous polarization as well as tensile strain and space charge effects due to ionized acceptors and free holes. In the figures, the energy zero is taken to be at the Fermi energy shown as a dotted black line and therefore the energy of the valence band is negative in sign. We shall discuss the energy in terms of magnitude from the Fermi level and so effectively reverse the sign of the energy shown in the figure in our discussion. Therefore moving into the energy gap is taken as negative in sign while increasing energy within the well away from the valence band edge is taken as increasing in energy. This allows us to compare the physics with that of an n-doped conduction band superlattice.

The figures indicate the acceptor energy, represented by a dotted line positioned within the energy gap below the valence band edge of the QW and QB materials. The bottom and top energies of the miniband are shown as a green and black line, respectively. Two kinds of SPS are illustrated in Fig.1. Fig.1a shows the HH superlattice and Fig.1b the LH superlattice of a superlattice composed of GaN/AlN 2MLs/2MLs, where the GaN is the well material and AlN the barrier material. The material offset is 0.85ev[26] and the resultant offset including the piezoelectric, spontaneous polarization, stain and any space charge effects is 0.51eV. Fig.1c shows the HH superlattice and Fig.1d the LH superlattice of a superlattice composed of GaN/Al$_{0.2}$Ga$_{0.8}$N 4MLs/4MLs where the GaN is the QW material and AlGaN the QB material. The offset due to the material difference is 0.18eV while that including the other factors is 0.15eV. The acceptor energy for AlN an Al$_{0.2}$Ga$_{0.8}$N are 517meV and 265meV, respectively.

\begin{figure}
\centering
\begin{subfigure}{0.49\textwidth}
    \includegraphics[width=\textwidth]{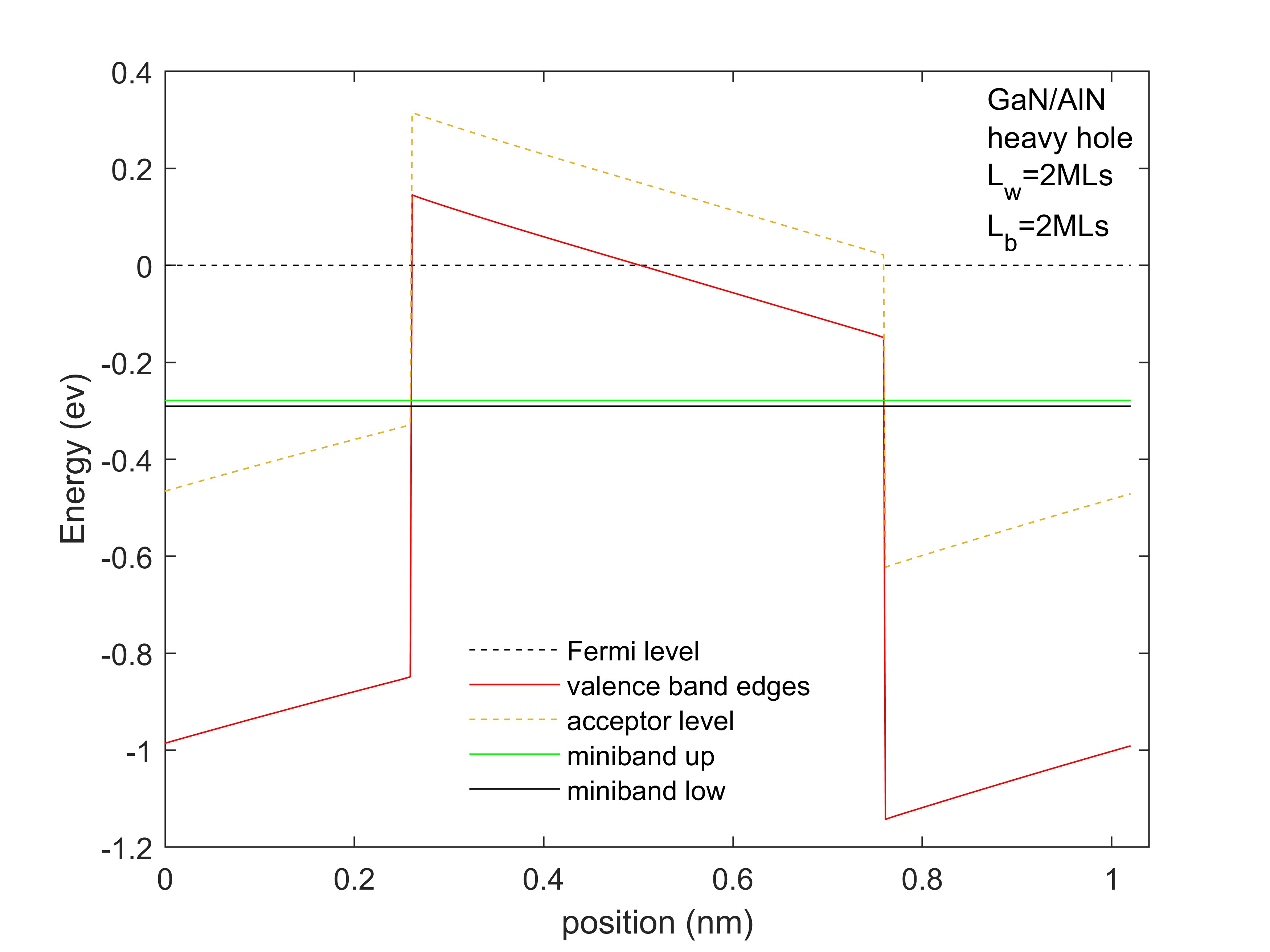}
    \caption{}
    \label{fig:first}
\end{subfigure}
\hfill
\begin{subfigure}{0.49\textwidth}
    \includegraphics[width=\textwidth]{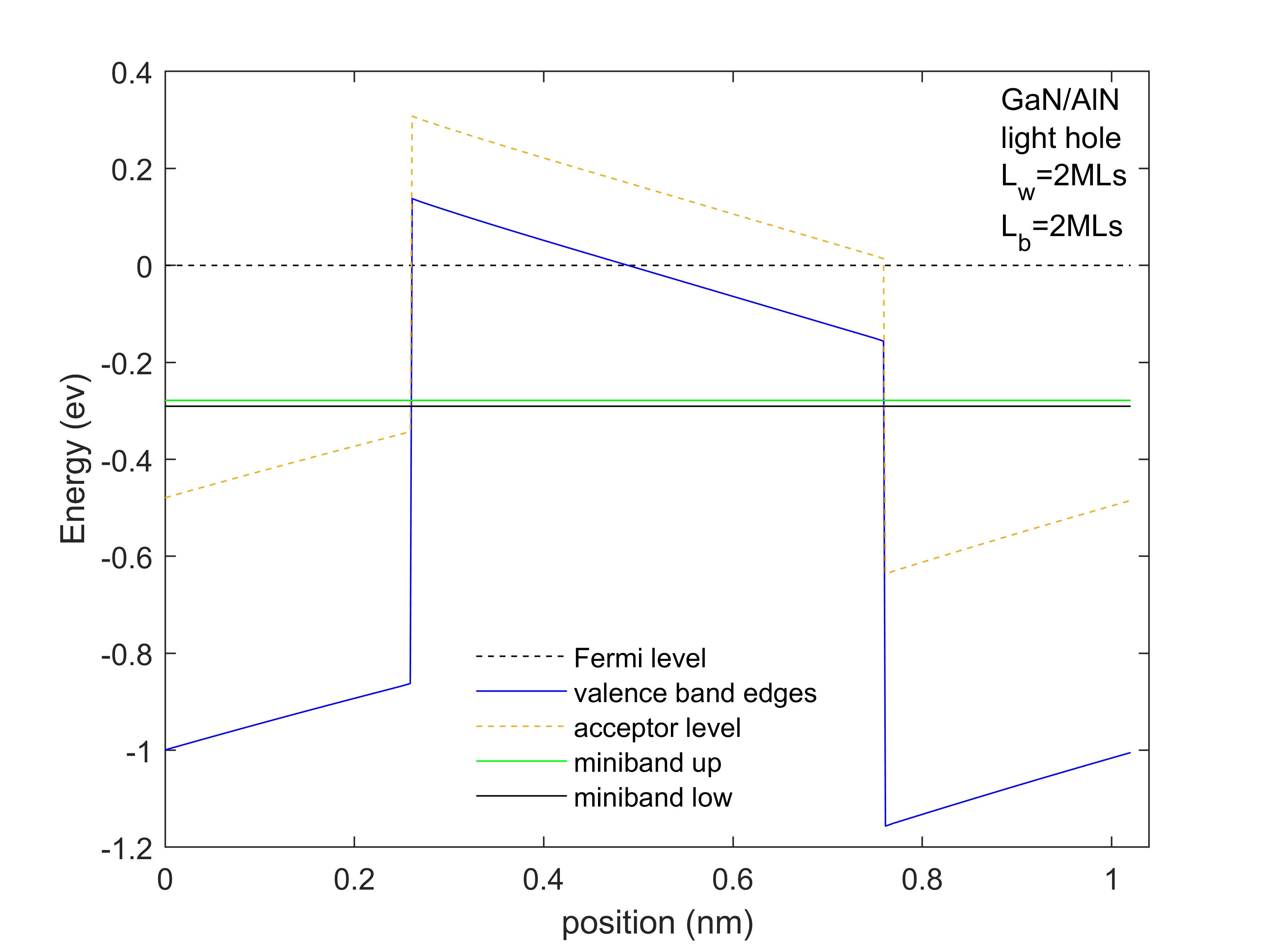}
    \caption{}
    \label{fig:second}
\end{subfigure}
\hfill
\begin{subfigure}{0.49\textwidth}
    \includegraphics[width=\textwidth]{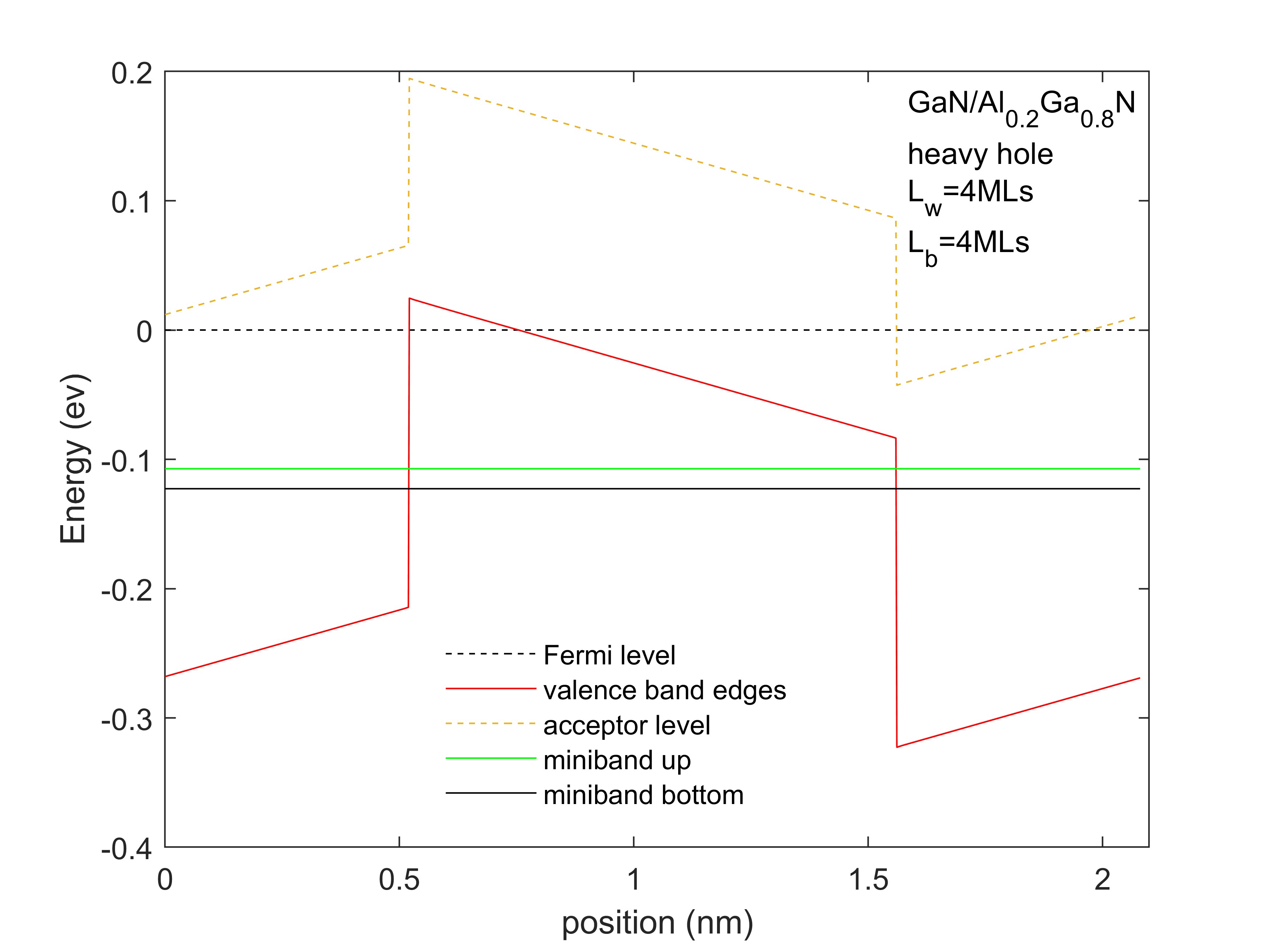}
    \caption{}
    \label{fig:first}
\end{subfigure}
\hfill
\begin{subfigure}{0.49\textwidth}
    \includegraphics[width=\textwidth]{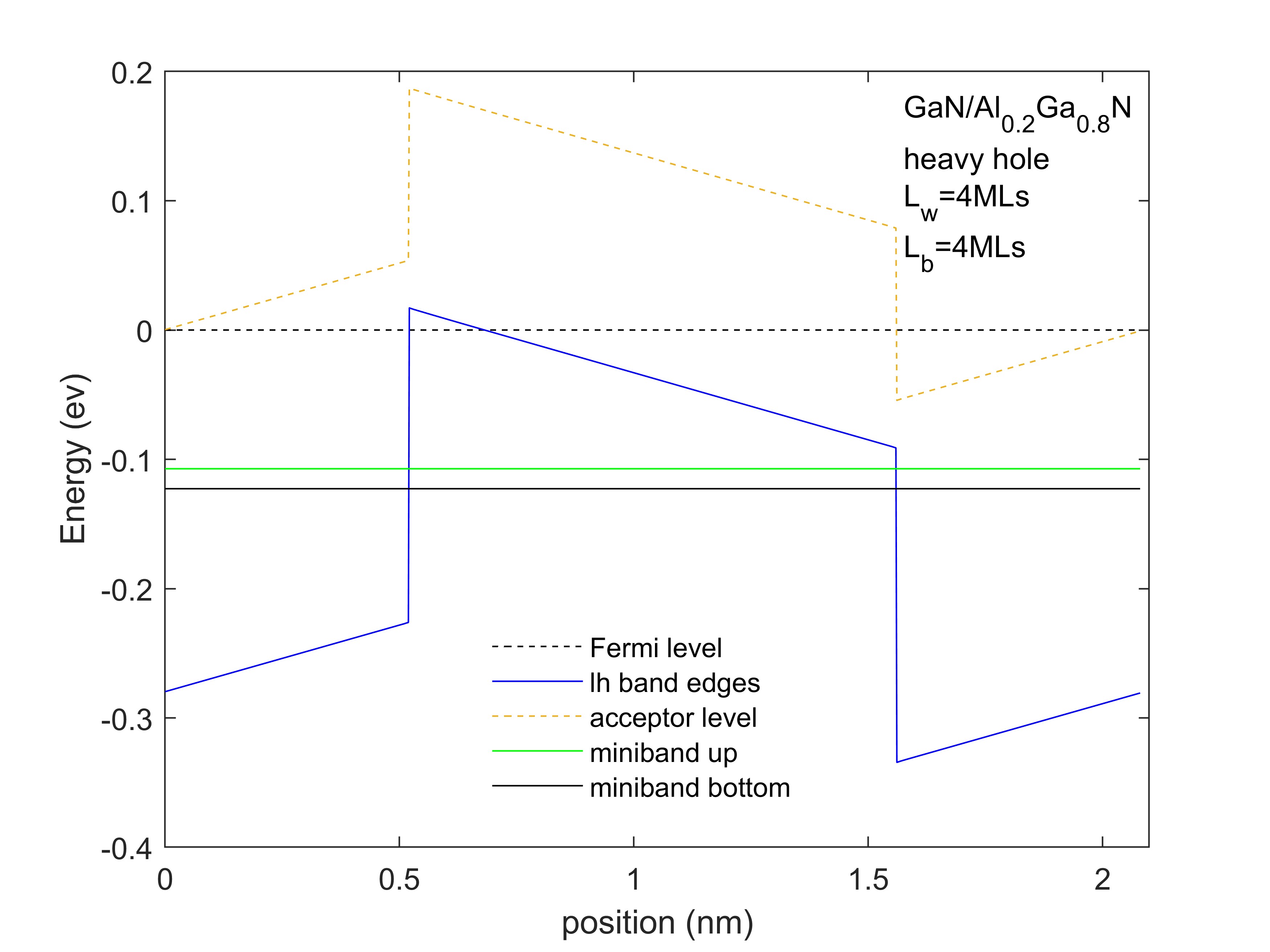}
    \caption{}
    \label{fig:second}
\end{subfigure}
\caption{Fig.1 shows the valence band edges, acceptor energy levels, Fermi levels, and miniband energy levels of the heavy hole (HH) (a,c) and light hole (LH) (b,d) minibands of: a) and b): a GaN/AlN 2MLs/2MLs superlattice: (c) and (d) a GaN/Al$_{0.2}$Ga$_{0.8}$N superlattice with 4MLs/4MLs. Both superlattice have N$_A$ = $10^{20}cm^{-3}$.}
\label{fig:figures}
\end{figure}

$L_{QW}$ and $L_{QB}$ were chosen to explore the position and energy thickness of the resulting minibands and to explore the band edge profiles and the acceptor ionization within the structures. The thickness chosen for the SPS with the AlN QB is very thin, demonstrating its large valence band offset in the valence band providing a high barrier minimizing the width of the miniband which even for 2MLS is very small. Although 2MLs should be possible to be grown whether this remains as a superlattice or an alloy is uncertain. Using band structure for 2MLs rather than atomistic modeling, which would be expected to be more appropriate, is hopefully indicative of trends rather than being numerically highly accurate. The band structure modeling used in this paper for SPS comprising 4MLs thicknesses should be producing results that are fairly accurate.

Fig.1 shows strong band bending due to the in-built piezoelectric, spontaneous polarization and tensile strain effects and space charge effects due to doping and it is almost 2$\times$ larger for the QW of half the size in the system with the higher QB of AlN than in the system with the Al$_{0.2}$Ga$_{0.8}$N QB. The acceptor depth is also much deeper into the energy gap for the AlN QB (517meV) than the Al$_{0.2}$Ga$_{0.8}$N QB (265meV). By examining the position of the Fermi energy with the energy position of the acceptor binding energy we can see if the acceptors are fully ionized or not. We expect using this modulation doping scheme that only the acceptors in the barrier will be fully ionized at all temperatures (as in Fig.1a and b) and transfer their holes into the QW and superlattice but observe that with these high built-in fields, some acceptors in the barriers will not be fully ionized at all temperatures (Fig.1c and d) but at room temperatures all acceptors in the QBs should transfer their holes into the QWs.

We have separated the miniband of the light and heavy holes ignoring band mixing at the interface of the QW and QB. Since the superlattice associated with the bulk pure HH and pure bulk LH are similar in position and energy width (varying less than 10meV) this suggests that this is not a significant approximation and results purely from their effective mass difference which is small in bulk. We therefore focus on the HH miniband but must remember to include scattering between the HH and LH in transport calculations or add together their joint density of states in transport considerations of a mixed HH and LH band.

We shall now consider the miniband position and width as a function of the QB material, $L_{QW}$ and $L_QB$. Fig.1 shows that even with a barrier width of only 2MLs the SPS with the AlN QB has a narrow miniband width of only 12 meV. In Fig.1c and 1d, where the Al concentration in the QB is reduced to 0.2 resulting in a smaller QB and increasing the QB width to 4MLs results in a wider miniband width of 15.7 meV. Although the acceptor energy level of the QB region also decreases with the low proportion of Al (around 250mev from AlN to Al$_{0.2}$Ga$_{0.8}$N), it is not as significant as the change in barrier height (around 700mev from AlN to Al$_{0.2}$Ga$_{0.8}$N). It is seen that the width of the miniband depends crucially on the QB height which depends on the Al content. The size of the QW determines the energy position of the miniband with a smaller QW pushing the energy level higher in energy relative to the valence band edges.
 
We also present in Fig.2 the position of the band edges and HH miniband without doping for the SPS shown in Fig.1 to examine, in particular, the Fermi energy (keep Fermi level at energy zero) and the space charge effects on the band edges. The shift of the band edges is expected and reflects that, without doping, no holes will be present in the valence band. The effect on the band edges allows us to separate space charge effects from the piezoelectric, spontaneous polarization and strain effects. we can see in Fig.2 that the SPS with AlN QB shows space charge effects in the QB region due to the large ionization of acceptors in the QB. Various doping concentrations, ranging from $10^{18}cm^{-3}$to $10^{22}cm^{-3}$, were also modeled but did not show much difference from the $10^{20}cm^{-3}$ presented in Fig.1. The observation of band edges bending in the figure is not readily discernible due to the short period nature of the SPS.

\begin{figure}
\centering
\begin{subfigure}{0.49\textwidth}
    \includegraphics[width=\textwidth]{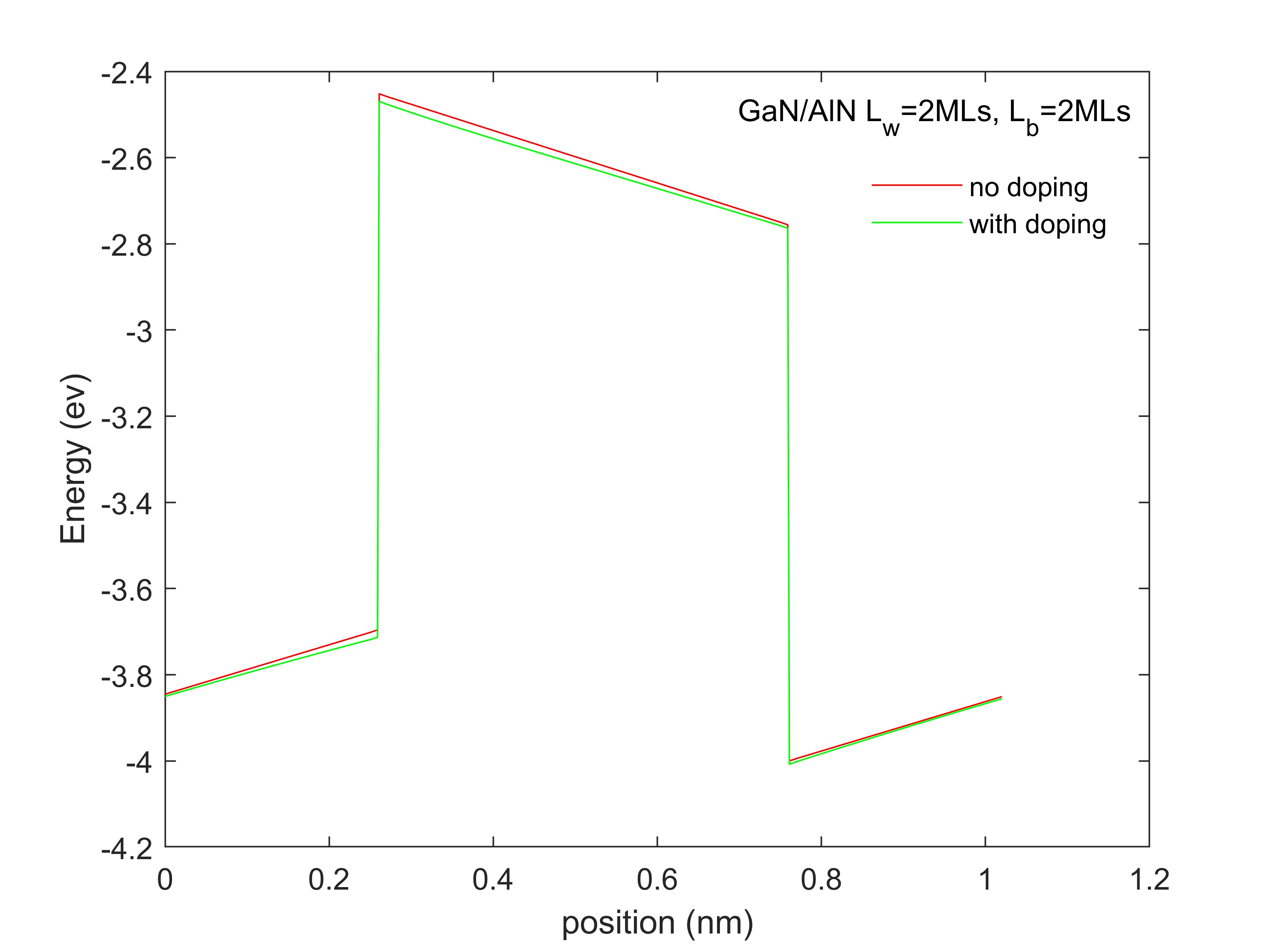}
    \caption{}
    \label{fig:first}
\end{subfigure}
\hfill
\begin{subfigure}{0.49\textwidth}
    \includegraphics[width=\textwidth]{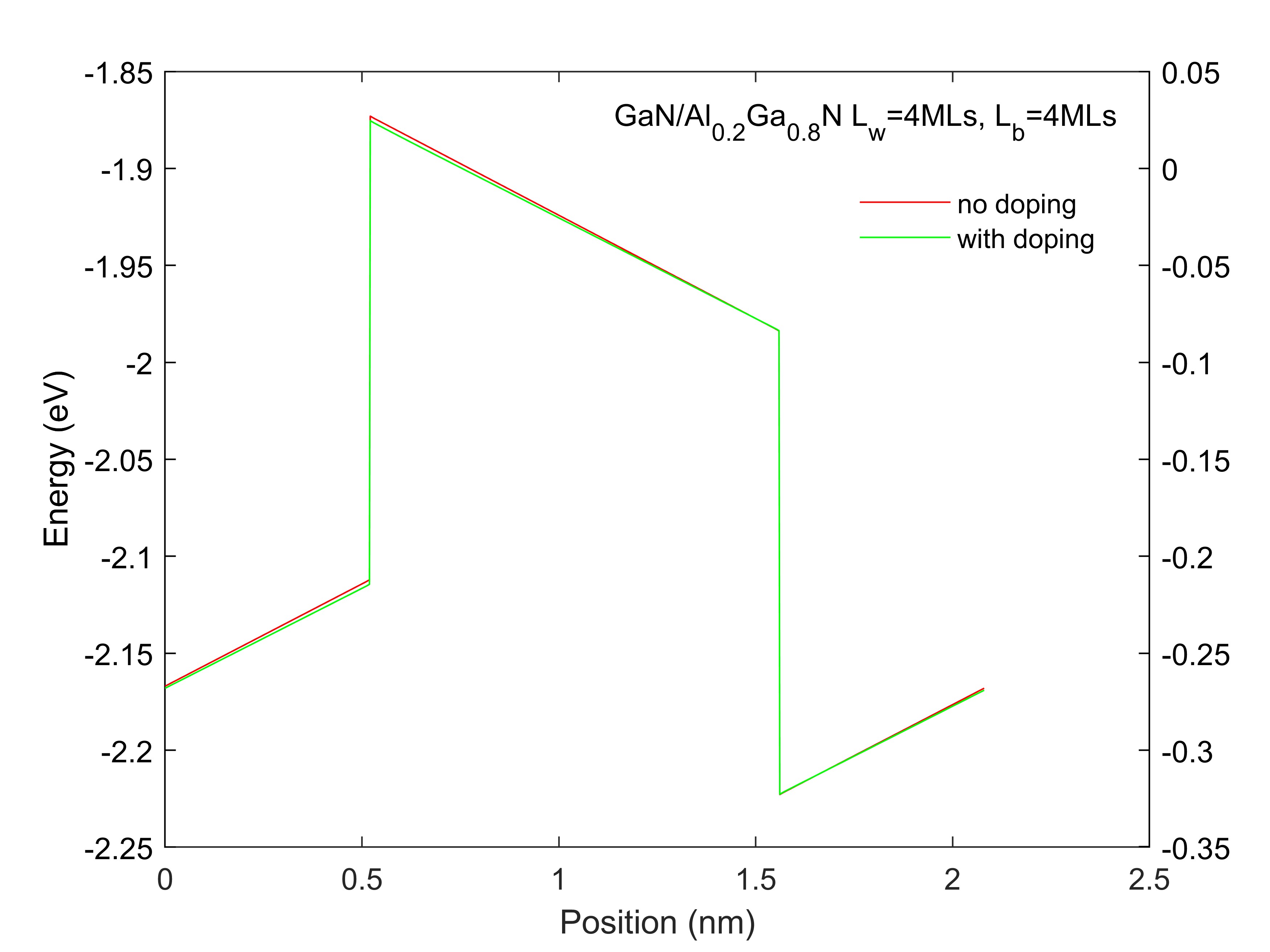}
    \caption{}
    \label{fig:second}
\end{subfigure}
\caption{(a), (b) shows the valence band edges associated with the HH superlattice of: a)  GaN/AlN $L_{QW}$=2MLs $L_{QB}$=2MLs and b) GaN/Al$_{0.2}$Ga$_{0.8}$N $L_{QW}$=4MLs $L_{QB}$=4MLs superlattice with no doping respectively. The red line shows the band edges without doping (use left y-axis) and the blue line (use right y-axis) shows them with doping N$_A$ = $10^{20}cm^{-3}$.}
\label{fig:figures}
\end{figure}

In Fig.3, we present an investigation into the energy position of the miniband energy and its width as a function of the QB composition (QB height) and $L_{QW}$ and $L_{QB}$. Unequal $L_{QW}$ and $L_{QB}$are considered. Fig.3a and Fig.3b show the energy of the first miniband energy level as a function of QW width and QB width, respectively. The solid and dashed curves depict the energy at the $\Gamma$ and X points showing the top and bottom energy of miniband in (0001) crystal direction. Distinct colors are utilized to represent materials with varying Al compositions. When the $L_{QB}$ is varied from 2MLs to 8MLs, the $L_{QW}$ is held constant at 4MLs; when the $L_{QW}$ is altered (also 2MLs to 8MLs), the $L_{QB}$ width is maintained at 4MLs.

\begin{figure}
\centering
\begin{subfigure}{0.49\textwidth}
    \includegraphics[width=\textwidth]{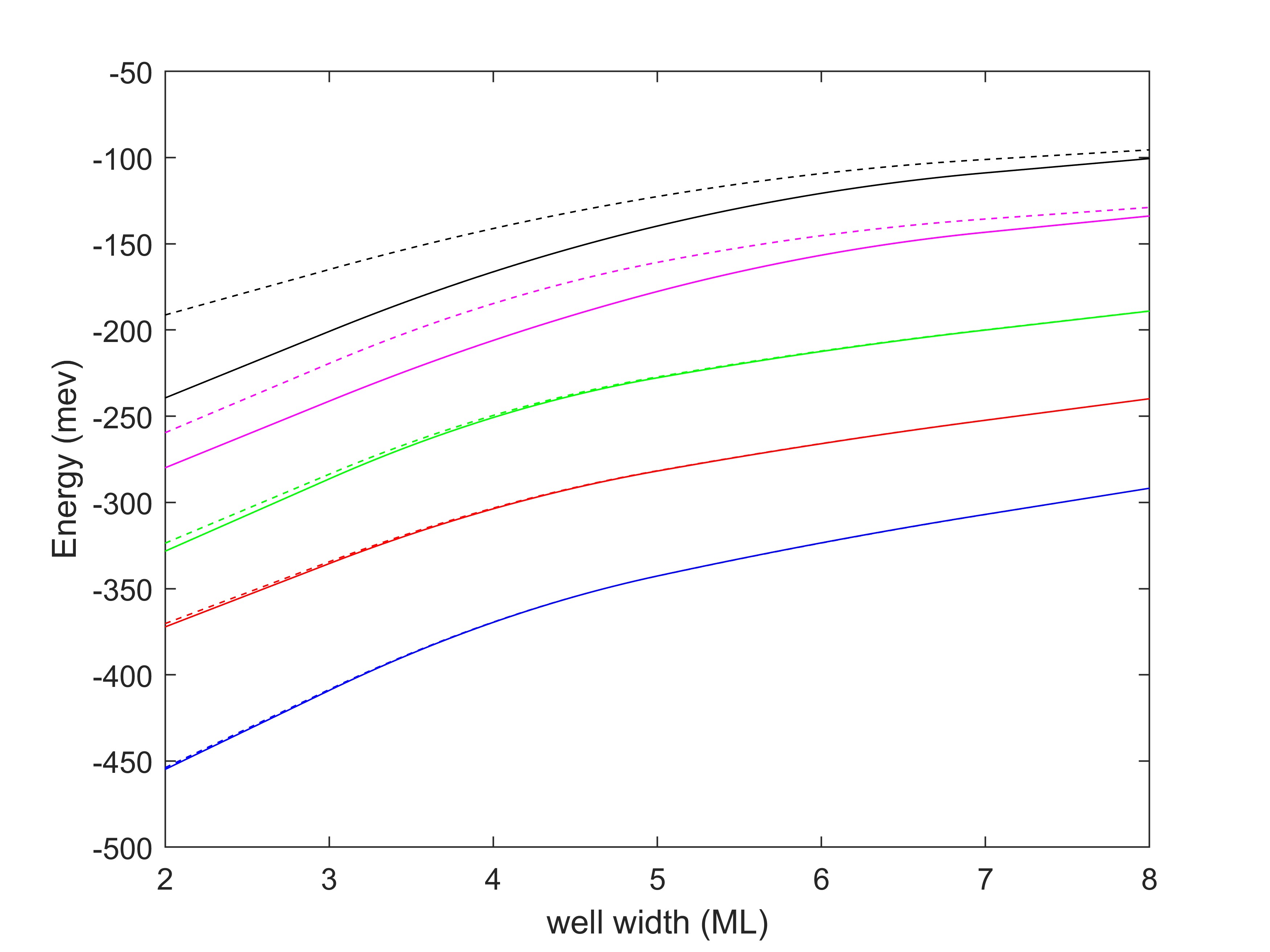}
    \caption{}
    \label{fig:first}
\end{subfigure}
\hfill
\begin{subfigure}{0.49\textwidth}
    \includegraphics[width=\textwidth]{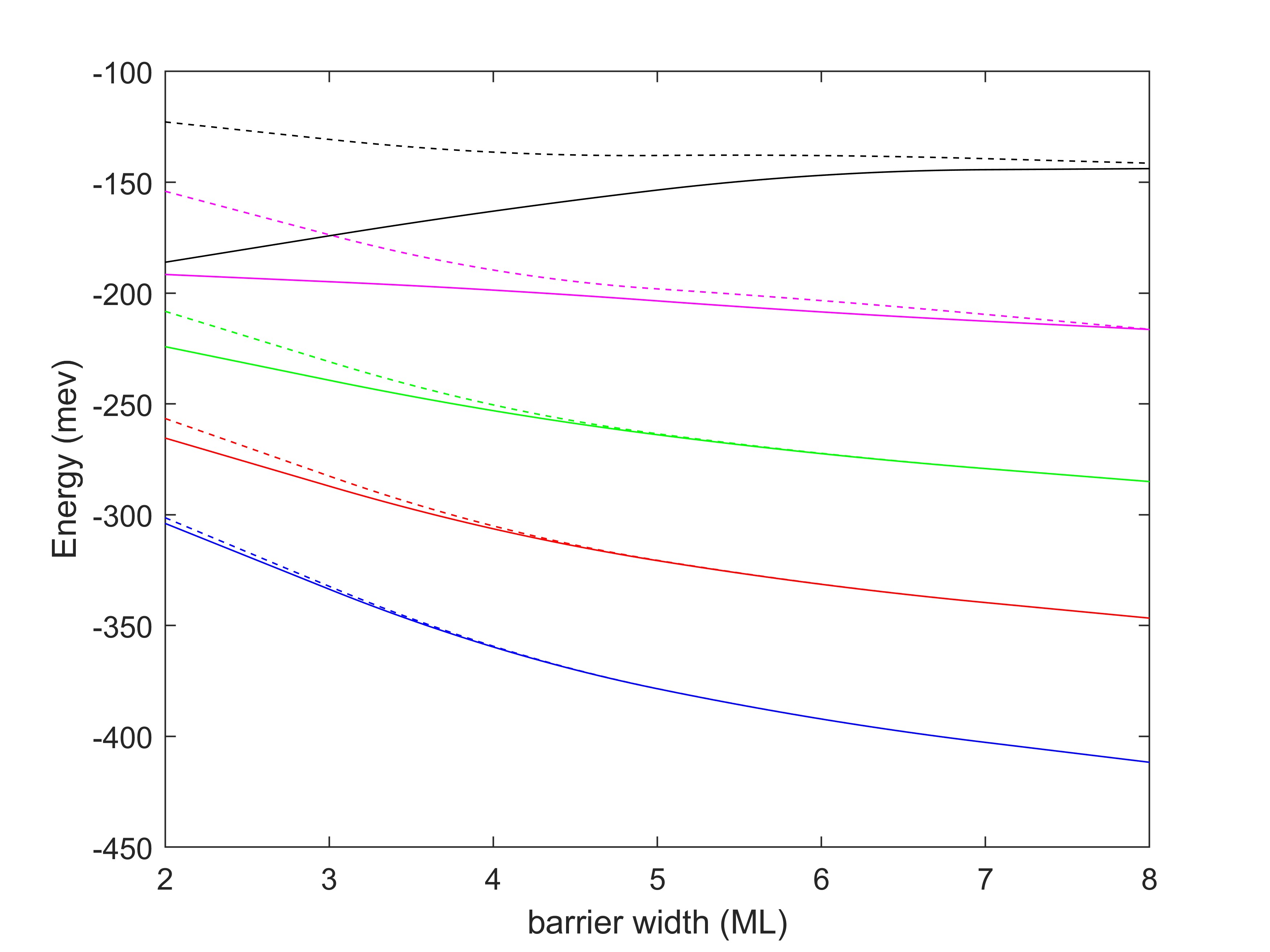}
    \caption{}
    \label{fig:second}
\end{subfigure}

\caption{Solid and dashed lines designate the top and bottom energy (respectively) of the superlattice miniband. Distinct colors represent various QB materials: blue for AlN, red for Al$_{0.8}$Ga$_{0.2}$N, green for Al$_{0.6}$Ga$_{0.4}$N, pink for Al$_{0.4}$Ga$_{0.6}$N, and black for Al$_{0.2}$Ga$_{0.8}$N. The impact of modifying the $L_{QW}$ while keeping the $L_{QB}$ at 4MLs is demonstrated in Fig.3a, whereas Fig.3b illustrates the impact of changing the $L_{QB}$ while keeping the $L_{QW}$ width at 4MLs.}
\label{fig:figures}
\end{figure}

The superlattices with a smaller proportion of Al have a wider miniband width are shown in Fig.3, consistent with the analysis in Fig.1. It can be observed that when the $L_{QW}$=$L_{QB}$, a superlattice with a low Al composition QB exhibits a more prominent miniband width. Furthermore, we discovered that increasing the size of either the QW or the QB can effectively reduce the miniband width. As the $L_{QW}$ or $L_{QB}$ increases, the miniband width gradually reduces until the energy levels of the upper and lower miniband limits coalesce to the energy level of the QW. However, it is also evident that the $L_{QB}$ has a greater influence on the miniband width than the $L_{QW}$. For instance, in GaN/AlN, as the QB increases from 2MLs to 4MLs, the changes of the miniband width are $\Delta$E=3.65meV, whereas an increase in QW width from 2MLs to 4MLs leads to a change of only $\Delta$E=0.95meV. The same trend is observed for superlattices with other Al compositions.

The energy level of the miniband also shifts due to changes in Al composition and size. Due to the greater height of the QB with higher Al composition, the energy level of the miniband will shift upwards from the valence band edge. When only the $L_{QW}$ is increased, the energy level of the miniband decreases; however, when the $L_{QB}$ increases (increasing well confinement), the energy level increases. It should be noted that for GaN/Al$_{0.2}$Ga$_{0.8}$N superlattices with low QB heights, the energy of the bottom of the miniband shows an upward trend as its width shrinks to approach the discrete energy of the QW. We have demonstrated that a miniband can be tuned through a range of energy positions and energy widths by choice of QB composition and $L_{QW}$ and $L_{QB}$.

We can calculate the longitudinal effective mass $m_v$ of the superlattice miniband by examining the hole miniband dispersion for the various superlattices. The dispersion of the minimum of the miniband energy against the momentum is cos-like with an inflection point in the middle of the band (between the gamma and x points) but at the gamma point, the dispersion can be fitted to a parabolic form and an effective mass assigned to the superlattice miniband. Large effective masses have low mobility and low effective masses have high mobility. The effective mass of the superlattice is determined by how much wavefunction probability resides in the QW and QB. If most of the wavefunction is in the QW the effective mass with be similar to that of the QW material while if more wavefunction resides in the QB the miniband effective mass will have an effective mass closer to that of the barrier material.

The effective masses of the miniband $m_v$ are shown in Fig.4 for $L_{QW}$=$L_{QB}$ for GaN/AlN and GaN/Al$_{0.2}$Ga$_{0.8}$N (with a few nonequal cases shown.) For small periods up to about 7 monolayers of $L_{QW}$=$L_{QB}$ layers, the effective mass does not vary much for the GaN/Al$_{0.2}$Ga$_{0.8}$N but then it increases above $m_v$>10$m_0$ while for GaN/AlN $m_v$ >10$m_0$ for any thickness >3 MLs for the $L_{QW}$=$L_{QB}$. This is different from reference[9] where because low effective masses were used for GaN and especially AlN much lower effective masses were found for the superlattice miniband. The GaN $m_v$=1.1$m_0$ while the AlN $m_v$=3.53$m_0$[26][27] so to keep the vertical mobilities high and keep substantial miniband energy width we want to consider low Al compositions for the QB. For GaN/Al$_{0.2}$Ga$_{0.8}$N 4MLs/4MLs the effective mass of the superlattice miniband is 2.65$m_0$ while it goes up to about 20$m_0$ for GaN/AlN 4MLs/4MLs. We will focus on studying GaN/Al$_{0.2}$Ga$_{0.8}$N superlattices with $L_{QW}$=$L_{QB}$<8MLs to keep the $m_v$ small. From here on we will just consider GaN/Al$_{0.2}$Ga$_{0.8}$N superlattices as having potentially the highest mobility.

\begin{figure}[ht]

\centering
\includegraphics[width=0.6\linewidth]{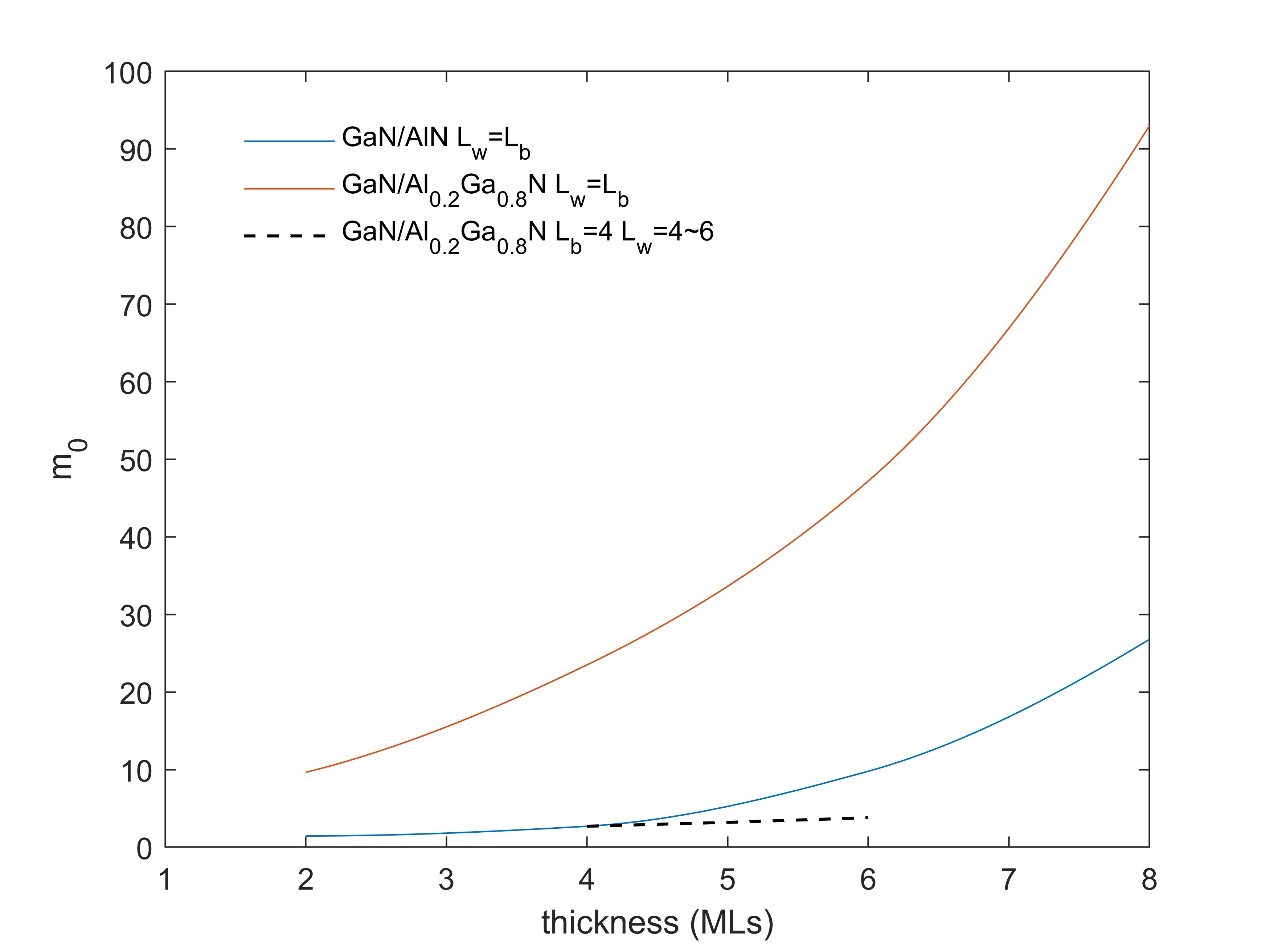}
\caption{Effective mass of heavy hole of GaN/AlN (red) and GaN/Al$_{0.2}$Ga$_{0.8}$N (blue), keep $L_{QW}$=$L_{QB}$. The effective mass for unequal $L_{QW}$ and $L_{QB}$ for GaN/Al$_{0.2}$Ga$_{0.8}$N (black) in few cases also shown.}
\label{fig:stream}
\end{figure}

In Fig.5, we present the free hole density in the superlattice miniband shown in Fig.1 (GaN/AlN 2MLs/2MLs and GaN/Al$_{0.2}$Ga$_{0.8}$N 4MLs/4MLs) showing the that the hole distribution is located chiefly in the QW and is adjacent the QB which donated the holes and where the ionized acceptors are located. The concentration of ionized acceptors is taken to be equal to that of the concentration of holes in the QW (charge neutrality). These representations enable the calculation of the average hole concentration, providing further insight into the characteristics of the system. The average concentration of free holes (averaged over the $L_{QW}$+$L_{QB}$) in GaN/AlN 2MLs/2MLs is 6.45$\times10^{18}cm^{-3}$ while the concentration of free holes in GaN/Al$_{0.2}$Ga$_{0.8}$N 4MLs/4MLs is 4.1$\times10^{17}cm^{-3}$. It is also possible to consider the calculation of the free holes in the QW which is relevant for lateral transport in the QW, for equal $L_{QW}$=$L_{QB}$ it is just 2$\times$ the average over the period. We observe that the average hole concentration in the range of 2-8 (MLs) for $L_{QW}$=$L_{QB}$ varies from 3.1$\times10^{17}cm^{-3}$ to 7.3$\times10^{17}cm^{-3}$. We then investigated free hole population by varying the QB holding the QW value fixed at 4MLs and then varied the QW whilst holding the QB at 4MLs for a superlattice of GaN/Al$_{0.2}$Ga$_{0.8}$N. In both cases, the average hole concentration increases as is shown in Fig.6. It can be seen that changing $L_{QW}$ or changing $L_{QB}$ has almost the same effect on the hole concentration, which will make it from about 3.2$\times10^{17}cm^{-3}$ to about 5.1$\times10^{17}cm^{-3}$. The increased in $L_{QB}$ donates more holes into the QW increasing the free hole concentration while increasing the size of the QW reduces the energy position of the superlattice, favoring more holes to localize there. The free hole concentration of the average alloy can be calculated for Al$_{0.1}$Ga$_{0.9}$N, corresponding to the average composition of the GaN/Al$_{0.2}$Ga$_{0.8}$N superlattice will be around 1.5$\times10^{17}cm^{-3}$ showing that the superlattice increases the free hole concentration about 3$\times$ over the average hole concentration. It is also important to note that the hole concentration in the superlattice will be temperature independent while that in the alloy will be very temperature dependent. The following equation shows the relation of conductivity $\sigma$, the mobility $\mu$ is inversely proportional to mass in direction of travel (mv or mxy), and the hole density p in the usual equation showing how increasing p should increase $\sigma$ in direct proportion.

\begin{figure}
\centering
\begin{subfigure}{0.49\textwidth}
    \includegraphics[width=\textwidth]{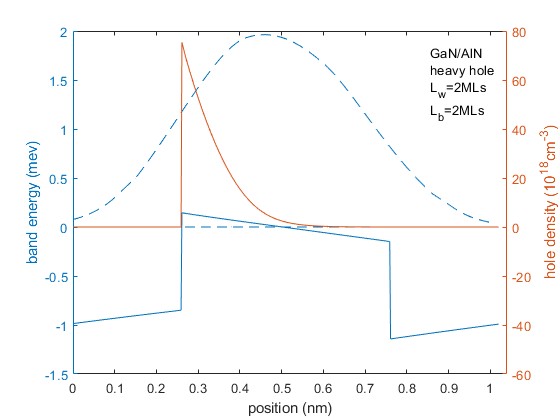}
    \caption{}
    \label{fig:first}
\end{subfigure}
\hfill
\begin{subfigure}{0.49\textwidth}
    \includegraphics[width=\textwidth]{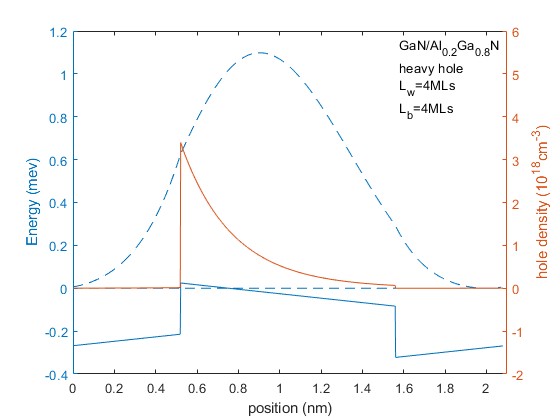}
    \caption{}
    \label{fig:second}
\end{subfigure}

\caption{(a) shows the band edges (blue solid line),  hole density (red solid line) of GaN/AlN 2MLs/2MLs, the left vertical axis is band energy, the right vertical axis is for hole density. (b) shows these for GaN/Al$_{0.2}$Ga$_{0.8}$N 4MLs/4MLs. In both figures wave function squared is shown  (blue dashed line) without scale.}
\label{fig:figures}
\end{figure}

\begin{figure}[ht]
\centering
\includegraphics[width=0.6\linewidth]{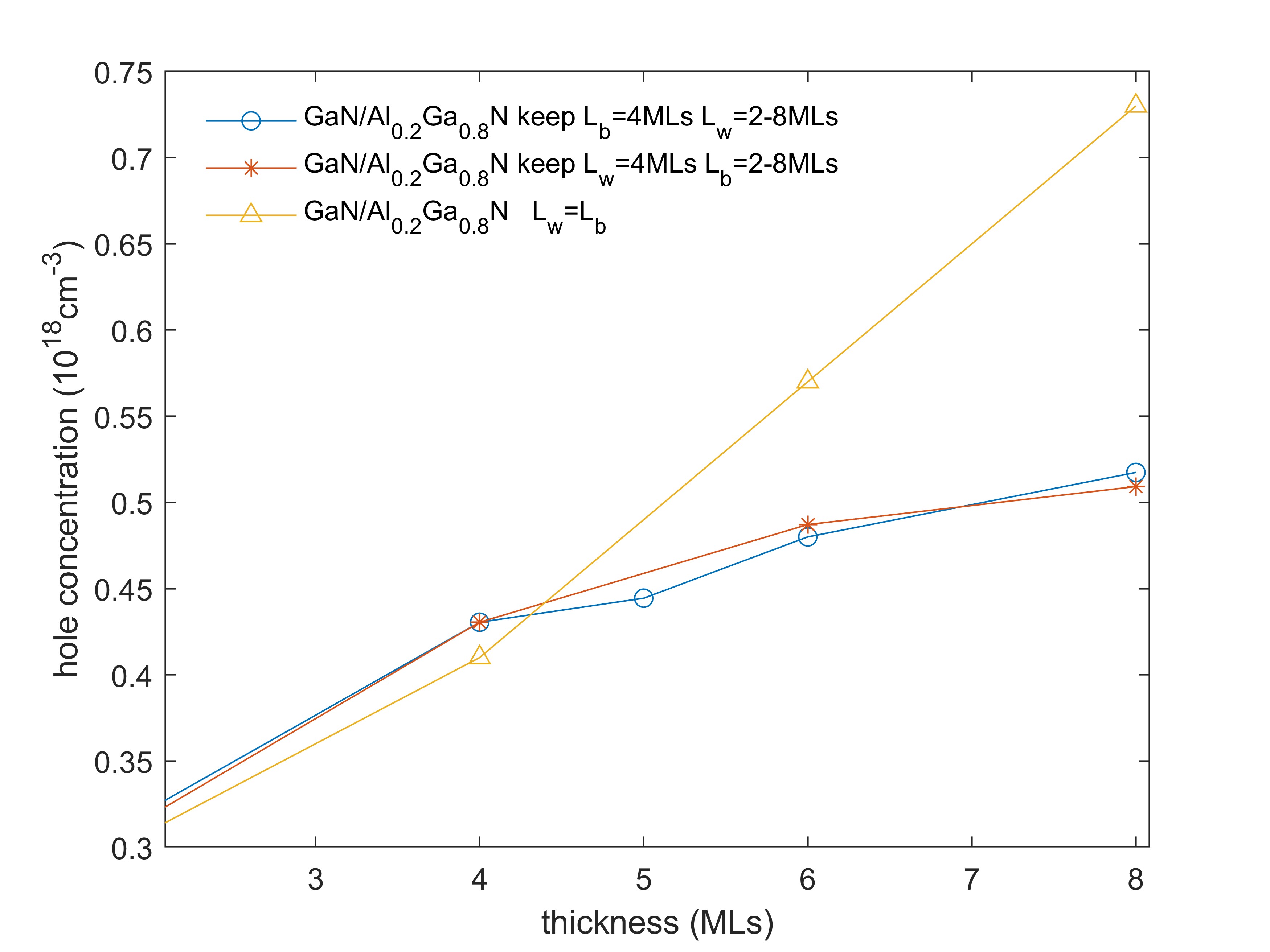}
\caption{The averaged hole density for an GaN/Al$_{0.2}$Ga$_{0.8}$N superlattice showing the case for $L_{QW}$=$L_{QB}$ (yellow line). The figure also shows the hole population by varying $L_{QB}$ holding the $L_{QW}$ value fixed at 4ML (blue line) and the hole population by varying $L_{QW}$ whilst holding the $L_{QB}$ at 4ML (red line).}
\label{fig:stream}
\end{figure}

 \begin{equation}
\sigma = p\times e\times \mu\label{7}\\
\end{equation}

Our trends agree well with Reference[9] in $m_v$ not varying much with varying equal $L_{QW}$ and $L_{QB}$ over the size range considered in this work (2-8MLs corresponding to 0.5-2nm)and showing enhancement of the concentration of free carrier holes in the superlattice. Reference[9] showed that it is possible to increase the hole average density by increasing the size of the $L_{QW}$=$L_{QB}$ but then we move into the region of a small energy width miniband towards multi quantum well (MQW) which will not offer good vertical transport. Thus the averaged resistivity which is inversely proportional to hole concentration is reduced by the same amount. The lateral conductivity is enhanced by the same amount assuming no additional scattering is introduced. This appears to be observed experimentally[30] but is much more observable for multi quantum barrier (MQB) rather than SPS. Theoretically[9] find an enhancement from 2$\times10^{17}cm^{-3}$ (in bulk film) to 6$\times10^{17}cm^{-3}$ an increase of about 3$\times$ using an GaN/Al$_{0.2}$Ga$_{0.8}$N 1nm/1nm superlattice doped 5$\times10^{19}cm^{-3}$. They find that as the $L_{QW}$=$L_{QB}$ increases, the hole concentration increases up to 5$\times10^{18}cm^{-3}$. Experimentally[30] they measured 4$\times10^{17}cm^{-3}$ in bulk and measured 2$\times10^{17}cm^{-3}$ at 8MLs/8MLs, going up to 2.5$\times10^{18}cm^{-3}$ for 28MLs(7nm). These are MQB-they attribute low values  to interface scattering.

Fig.5 also shows the normalized wavefunction squared of the SPS miniband which has significant magnitude chiefly in the QW but shows clearly the occupation distribution in the QB. We see for the GaN/AlN 2MLs/2MLs superlattice that there is a similar wavefunction overlap into the QB as there is for the GaN/Al$_{0.2}$Ga$_{0.8}$N 4MLs/4MLs superlattice. This reflects that their miniband widths are similar as observed earlier. If the GaN /AlN superlattice is increased to 4MLs thickness for the QW and QB the wavefunction in the barrier reduces significantly.

\newpage

\subsection*{Vertical hole transport in the minibands}

Vertical Transport of the holes depends on the electric field experienced by the holes, the effective mass in the vertical dimension of the miniband $m_v$, and the various scattering mechanisms that they undergo. The value of the E field applied across the period must be considered relative to the energy width of the miniband to avoid the break-down of the miniband into an of Wannier-Stark ladder of isolated states in which vertical transport would be low. The criterion for the onset of the splitting of the miniband is $\delta$E/(q*L) where $\delta$E is the width of the miniband and L is the period. (For a miniband width of 15 meV and L=2nm ($L_{QW}$=$L_{QB}$=1nm) this corresponds to about 50 kV/m. The $m_v$ of the HH and LH minibands depends on the miniband structure and ranges with Al composition x see Fig.5. It was seen that the composition of Al in the QB must be low to keep $m_v$ small. The hole in the miniband will be accelerated by the resultant electric field in the z vertical direction that it experiences and the acceleration is inversely proportional to its $m_v$. Next, the scattering processes that the holes suffer traveling in the miniband need to be considered. Scattering between the LH and HH minibands needs to be considered but as the minibands overlap well should not pose a serious problem. We assume that the simple scattering process of acoustic phonon scattering, which is relatively small in terms of the total scattering rates should not be too detrimental. The largest scattering rate is h-LO phonon emission for holes accelerated high in the miniband with energy higher than the LO phonon energy (50mev) which will not be the case here as the minibands are much narrower in energy width-suppressing this scattering. Although the energy of the hole is reduced by the LO phonon energy its direction of travel in the miniband should not be deflected much from its direction of travel (in the direction of the applied field) due to the small angle of scatter of h-LO phonon scattering[31]. The acceptors in the AlGaN system are very deep in the bandgap in this system. Combined with the in-built E fields can bring the energy position of these II scatters close to the energy states of the holes in the miniband. (The holes in the miniband have a potential energy of zero at the lower miniband edge and gain in kinetic energy as they increase in energy in the miniband but are restricted to be in energy states within the miniband). The II scattering process is important in transport as although it is elastic (so no energy is lost from the hole) it significantly changes the momentum (direction of travel) of the holes, which has a large effect on the transport of the holes.

The calculation of the scattering of holes in the miniband by ionized impurities in the barriers is difficult to calculate. For bulk materials, the scattering of holes from ionized impurities can be related to whether
the hole has sufficient kinetic energy to be scattered by the Coulomb potential or becomes trapped by the Coulomb potential. The important parameter is the ratio of the Coulomb potential to the kinetic energy of the hole. This approach considers the hole to be in the same region of space and free to move in energy states around the ion. In the case of the miniband, the hole is confined in the miniband energy band and the II is at a fixed energy which can be used to restrict their interaction and reduce it below the Conwell Weiss II scattering rate[31]. Another consideration is the physical proximity of the hole to the II center. This can be considered in terms of the time through which the hole travels near the ion and feels its potential relative to the time it travels in the miniband not feeling the potential of the ion (being physically removed in space from the ion). This is very difficult to calculate as the Coulomb potential is long-range and will be partially screened by holes in the miniband. The fact that the ions are spatially located in the barrier can be used to reduce II scattering if the wavefunctions of the holes in the miniband can be tailored to have lower probability in the barrier where the ions are which is equivalent to them spending less time in the QBs. To study the II scattering rate we first calculate the Conwell Weiss II scattering in bulk and then consider how to tailor the wavefunction to include the effects described above that will minimize the II scattering.

We simplify the Conwell Weiss approach to calculate the maximum scattering rate of the hole with kinetic energy (taken to be 3/2 kT) equal to the Coulomb potential energy $q^2/(4\pi\varepsilon_sr)$, where r is the distance between the hole and ion which is the lowest energy hole that can be scattered rather than trapped by the Coulomb potential[32]. We can use this to calculate r and the $\pi r^2$-the scattering cross section which is $\sigma=\frac{q^4\pi}{(6\pi\varepsilon_skT)^2}$. From this, we can calculate the scattering time $\tau_{sc-C}$(time between the scattering events) according to the equation, in which $N_{sc-impurity}$ is the II dopant concentration (equal to the hole concentration) and $v_{th}$ is the thermal velocity of the hole.

 \begin{equation}
\tau_{sc-C} =\frac{1}{\sigma_{sc-C}v_{th}N_{sc-impurity}}\label{7}\\
\end{equation}

We use this to calculate the scattering rate (scattering events per unit time). From this, we can calculate the drift velocity according to:

 \begin{equation}
v_d=-\frac{q\tau_{sc-C}}{m^*}E\label{5a}
\end{equation}
where
 \begin{equation}
\mu_n=\frac{q\tau_{sc-C}}{m^*}\label{5b}
\end{equation}

The E is the applied electron field which is 2kV/cm here. In Table 2 results of this simplified scattering model are shown for a range of GaN/Al$_{0.2}$Ga$_{0.8}$N superlattices. The superlattices are not very different in thicknesses so their effective masses $m_v$ and hole concentrations are similar but we see that the mobility decreases with increasing $m_v$.

\begin{table}[ht]
\centering
\begin{tabular}{|l|l|l|l|}
\hline
  & $L_{QW}$=4MLs $L_{QB}$=4MLs & $L_{QW}$=5MLs $L_{QB}$=4MLs & $L_{QW}$=6MLs $L_{QB}$=4MLs\\
\hline
Effective mass & 2.65$m_0$ & 3.2$m_0$ & 3.8$m_0$\\
\hline
Hole concentration & 4.1$\times10^{17}cm^{-3}$ & 4.4$\times10^{17}cm^{-3}$ & 4.6$\times10^{17}cm^{-3}$\\
\hline
$v_{th}$  &7.24$\times10^{4}m/s$ &6.63$\times10^{4}m/s$  &6$\times10^{4}m/s$\\
\hline
Miniband width  & 15.7mev  & 10.1mev   & 7mev\\
\hline
Cross section   & 5.47$\times10^{-13}m^{2}$  & 5.47$\times10^{-13}m^{2}$   & 5.47$\times10^{-13}m^{2}$\\
\hline
Scattering time & 6.27$\times10^{-13}s$  & 6.36$\times10^{-13}s$  & 6.61$\times10^{-13}s$\\
\hline
Drift velocity  &8.1$\times10^{4}m/s$  &6.9$\times10^{4}m/s$  &6.1$\times10^{4}m/s$\\
\hline
Mobility       &405$\times cm^2/Vs$    &354$\times cm^2/Vs$   &305$\times cm^2/Vs$\\
\hline
\end{tabular}
\caption{\label{tab:example}The tensile strain of the QB under different Al compositions}
\end{table}

The Conwell-Weiss expression for the scattering rate is shown in Equation 11[31].

 \begin{subequations}
\begin{align}
\sigma_m=&2\pi(\frac{\mu}{k})^2{\log[(1+(\frac{1}{2}N_I^{-(1/3)})^2{(k/\mu)^2}]}\label{5a}\\
\mu=&Z(R^*_H/E_k)^{1/2}\label{5b}
\end{align}
\end{subequations}

The momentum relaxation cross-section, described in the Conwell Weiss approach relates the ionised impurity $R^*_H$ (the effective Rydberg energy of the acceptor) where Z is the charge on the acceptor to the kinetic energy of the free hole $E_k$. This ratio is defined as $\mu$ and the momentum relaxation cross section is proportional to the square of this ratio. The k corresponds to the momentum associated with the free hole kinetic energy. $N_{I}$ is the concentration of ionized impurities. This equation is thus similar to the simple analysis that we used.

We can conclude that to minimize scattering we want to maximize the difference in energy between the acceptor ion and the kinetic energy of the holes in the miniband. Therefore the energy states of the miniband should be far apart in energy from the energy of the ionized acceptors. At a minimum, the acceptor energy level should not reside within the miniband. For the GaN/Al$_{0.2}$Ga$_{0.8}$N 4MLs/4MLs, SPS there is no overlap, and the miniband energy level is far away from the acceptor level. Reducing the proportion of Al in the barrier material Al$_{x}$Ga$_{1-x}$N will increase the miniband width and keep it far from the upper acceptor energy level.

We shall now focus on minimizing the II scattering by controlling the miniband wavefunction in the superlattice design using the arguments described above. We anticipate that designs that minimize the probability of the hole being in the QB region (equivalent to minimizing the wave function squared in the barrier) will minimize II scattering. We therefore present the wavefunction squared in the QB region for various designs. The normalization condition for the region of a single superlattice period, which comprises $L_{QW}$+$L_{QB}$, should be 1. By computing the interval of the barrier, the extent to which II scattering may occur in this region can be determined. Since a single period consists of two half-barrier regions and a well region, we can calculate the area of the shaded region to obtain the probability density of the hole in the potential barrier region as shown in Fig.7. Fig.7 displays the valence bandages and wavefunction squared values of the $L_{QW}$=2MLs $L_{QB}$=2MLs GaN/AlN and $L_{QW}$=4MLs$L_{QB}$=4MLs GaN/Al$_{0.2}$Ga$_{0.8}$N superlattices. The shaded area represents the proportion of the barrier region, which is 18.5\% and 13.8\%, respectively. This value is also influenced by the change of the well depth due to the variation in Al composition and superlattice period.

\begin{figure}
\centering
\begin{subfigure}{0.49\textwidth}
    \includegraphics[width=\textwidth]{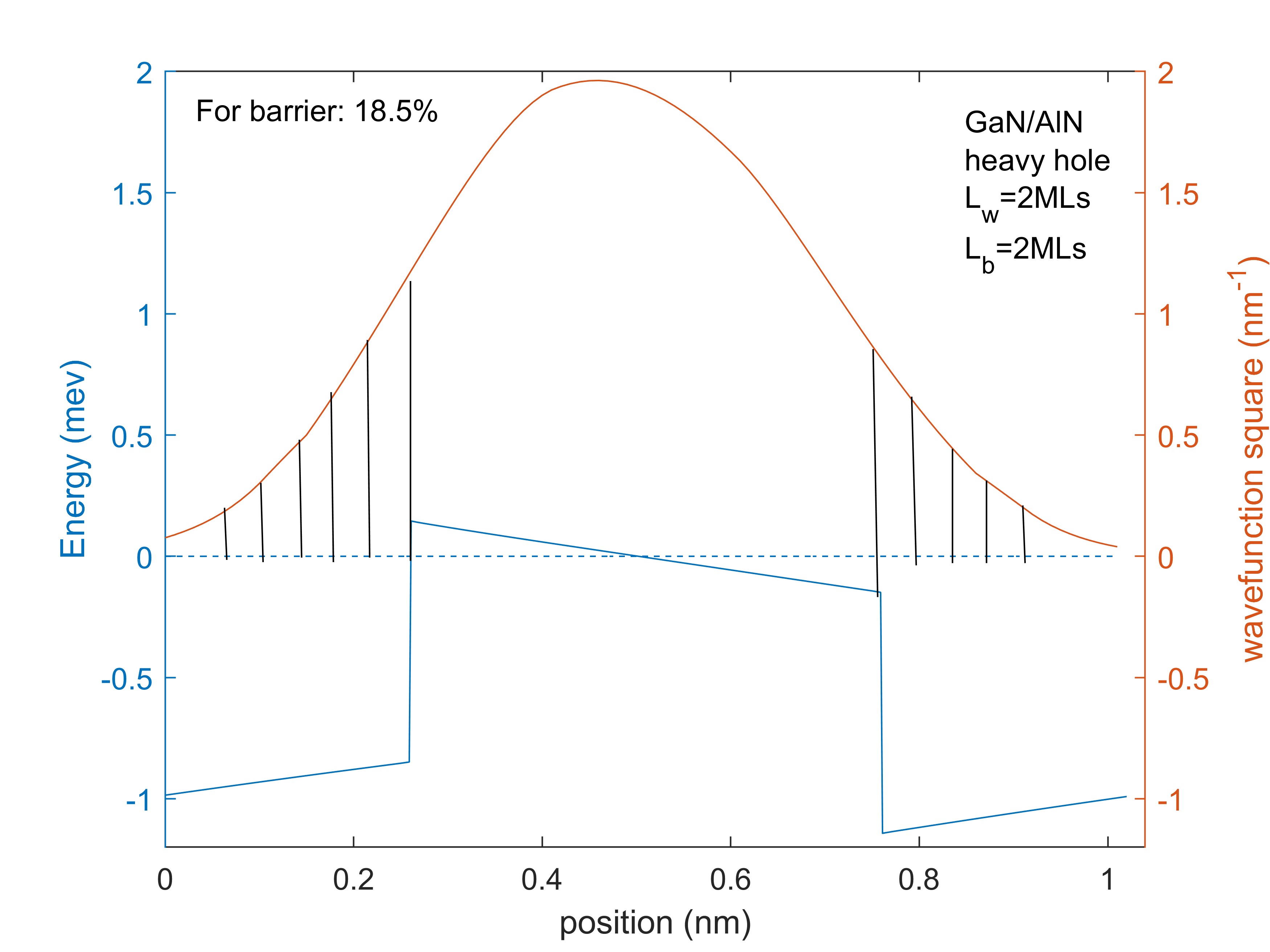}
    \caption{}
    \label{fig:first}
\end{subfigure}
\hfill
\begin{subfigure}{0.49\textwidth}
    \includegraphics[width=\textwidth]{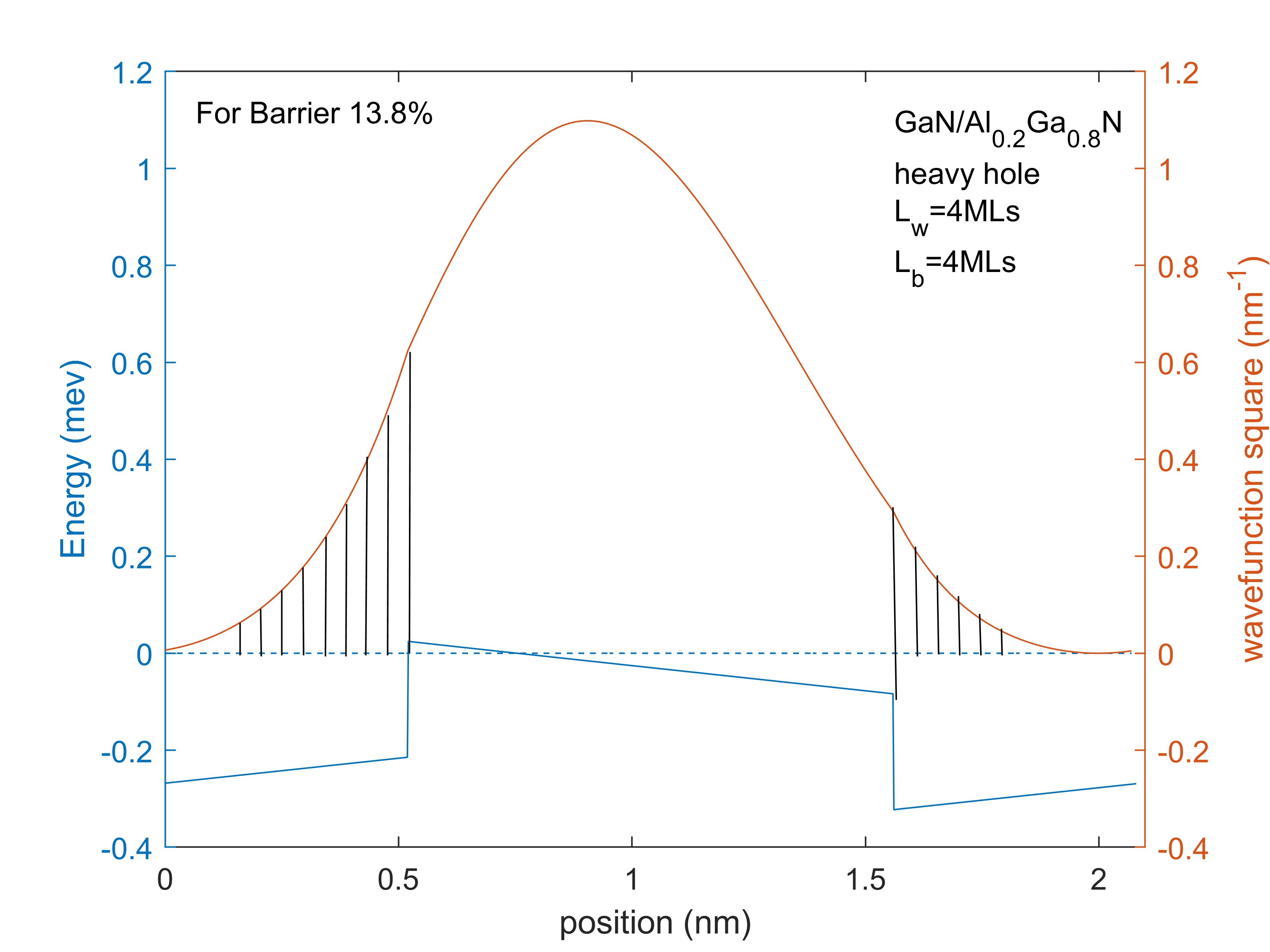}
    \caption{}
    \label{fig:second}
\end{subfigure}
\caption{(a), (b) represent valence band edges(blue) and wavefunction square(red) of GaN/AlN 2MLs/2MLs, GaN/Al$_{0.2}$Ga$_{0.8}$N 4MLs/4MLs respectively. Valence band edges can clearly give a period. The shaded area is the wave function squared in the barrier region and consists of two parts on either side of the QW.}
\label{fig:figures}
\end{figure}

It can be observed that the probability density of holes in the QB region of a periodic superstructure can be managed by adjusting the length of the QB and QW as seen in Fig.7. Reducing the thickness of the QB or increasing the thickness of the QW can reduce the probability density value for the same QW. As a result, structures that followed this design will experience fewer scattering effects from ionized impurities. Given these considerations, the $L_{QB}$ can be established as the minimum size required to construct a superlattice, which is 4MLs. On the other hand, the QW width can be expanded slightly to achieve a lower probability density in the barrier region. For example, if the GaN/Al$_{0.2}$Ga$_{0.8}$N superlattice has $L_{QW}$=2MLs and $L_{QB}$=2MLs, the probability density in the barrier region is  17.4\%. When $L_{QW}$=4MLs and $L_{QB}$=2MLs, the probability density decreases to 6.4\%, while $L_{QW}$=2MLs and $L_{QB}$=4MLs increase to  31.7\%. By adjusting the $L_{QW}$ and $L_{QB}$, the probability density can be more than doubled, which can significantly affect the scattering of ionized impurities. It should be noted that the GaN/Al$_{0.2}$Ga$_{0.8}$N superlattice with $L_{QW}$=4MLs $L_{QB}$=4MLs has a probability density of 13.8\%, which is different from the value of $L_{QW}$=2MLs $L_{QB}$=2MLs, even though they have the same barrier width ratio. This demonstrates that the probability density of the QB region is not directly proportional to the width of the QB in the period. Nevertheless, the approach to decrease this value remains the same, i.e., increase the QW width and reduce the QB width. Calculations were performed for different Al compositions, and the results were in agreement with the description provided above.

\subsection*{Conclusions}

There are 3 major criteria for designing a hole transporting miniband in the valence band of a GaN/AlGaN superlattice: 1) creating a miniband with sufficient energy width to support variations in the E field and any growth or interface fluctuations along the superlattice thereby creating a robust conducting channel for holes perpendicular to the superlattice, 2) minimizing the probability of finding the hole within the miniband in the QB (minimizing the wavefunction squared in the QB) and 3) maximizing the energy separation of the miniband from the ionized acceptors in the QB (particularly no overlap of the wavefunction energy with the ionized acceptor levels). To reduce II scattering, we can adjust the probability density of the hole in the QB by decreasing $L_{QB}$ and increasing $L_{QW}$. However, this design will shift the position of the miniband upward, as shown in Fig.3, causing the upper energy level to overlap with the acceptor energy levels. Therefore, considering the trade-off between these two conditions, a structurally-balanced design with minimal scattering theoretically exists. However, the weighted influence of these two factors on the results is still uncertain. A realistic superlattice with starts from a minimum QW and QB size of 4MLs. 

The design process starts with the criterion to reduce the overlap of the miniband with the ionized impurities in the QB. We start with a narrow QB and slowly increase the $L_{QW}$ to adjust the energy position of the miniband relative to the ionized impurities in the QB. $L_{QW}$ and $L_{QB}$ at this point are considered a potential design with the minimum scattering rate of ionized impurities now being considered. As the step size of the $L_{QW}$ is 1MLs (0.26nm), it is relatively easy to determine when the energy position of the ionized impurities and the miniband overlap. For high Al content, this is difficult to achieve. In the case of the GaN/AlN superlattice presented in Fig.1a, even when the $L_{QW}$ and $L_{QB}$ have been reduced to 2 MLs, an energy level overlap still occurs. Also, this superlattice structure has a very narrow energy width making it not a useful transporting channel. When the proportion of Al is reduced to 0.8 and 0.6, the overlap is less severe, although still present. It is not until the Al content drops to 0.4 that a non-overlapping structure is achieved with $L_{QW}$=$L_{QB}$=2MLs (Fig.8a). However, the QB width is still 2MLs, smaller than the minimum size to be considered (4MLs). When the proportion of Al is reduced to 0.2, the operability of the QW increases. As shown in Fig.1c and 1d, when $L_{QW}$=$L_{QB}$=4MLs, there is no overlap of the ionized impurities and the miniband, and the $L_{QW}$ can be further increased to decrease the QB probability of the hole to be in the QB, until it exceeds 6MLs, at which point the overlap will not occur. Fig.8b shows the GaN/Al$_{0.2}$Ga$_{0.8}$N superlattice structure with $L_{QW}$=5MLs and $L_{QB}$=4MLs has a 10.1meV miniband width and a 9.7\% probability of the hole being in the QB. We also texted for $L_{QW}$=6MLs, $L_{QB}$=4MLs GaN/Al$_{0.2}$Ga$_{0.8}$N superlattice. At this point, the probability of the hole in the QB region is 7.7\%, which is an improvement over the former case. However, in this case, the miniband width has been decreased to 6.7 meV because of the thicker QB. Compared to this, the two cases shown in Fig.8 may be better designs

\begin{figure}
\centering

\begin{subfigure}{0.49\textwidth}
    \includegraphics[width=\textwidth]{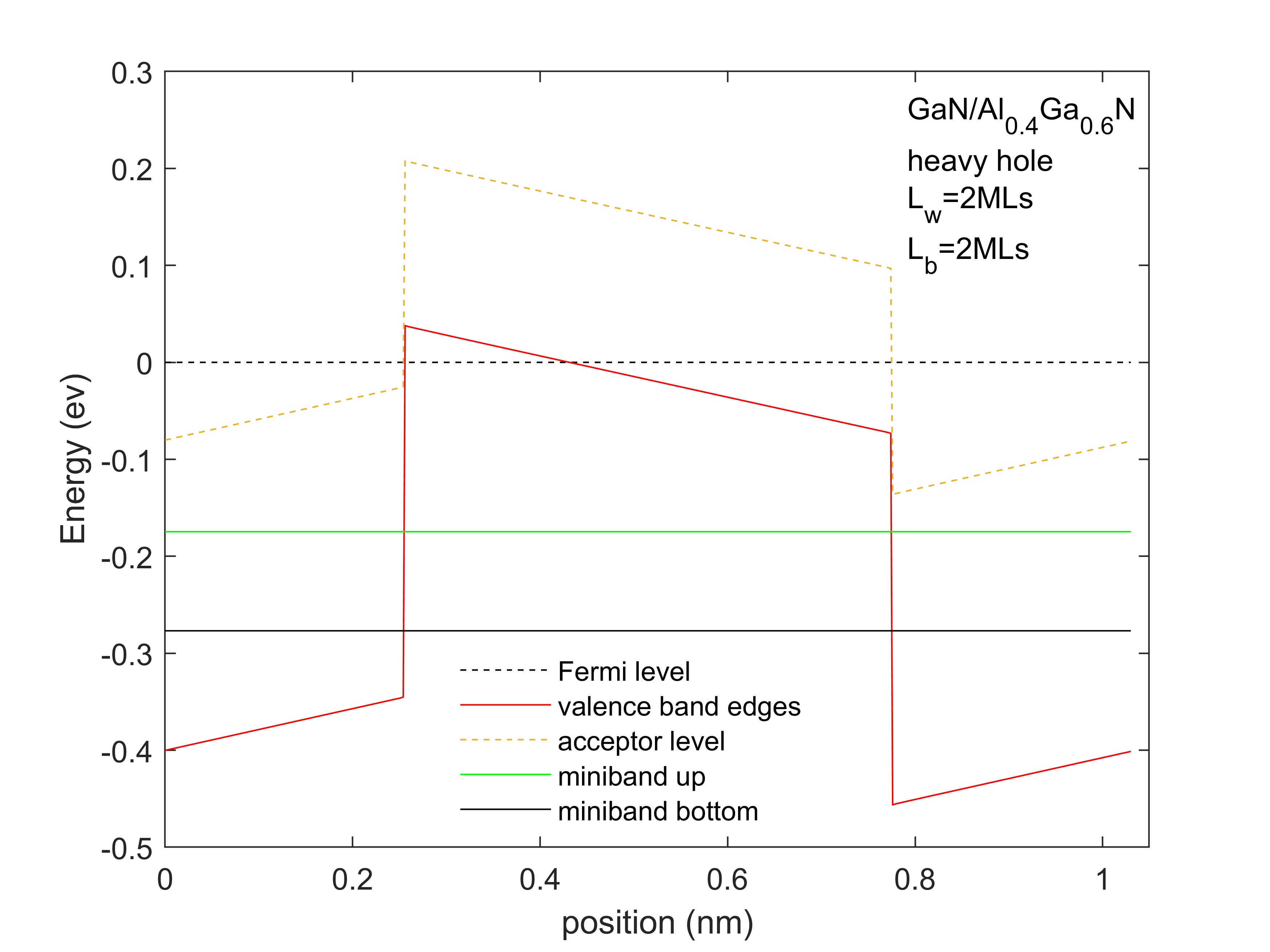}
    \caption{}
    \label{fig:second}
\end{subfigure}
\hfill
\begin{subfigure}{0.49\textwidth}
    \includegraphics[width=\textwidth]{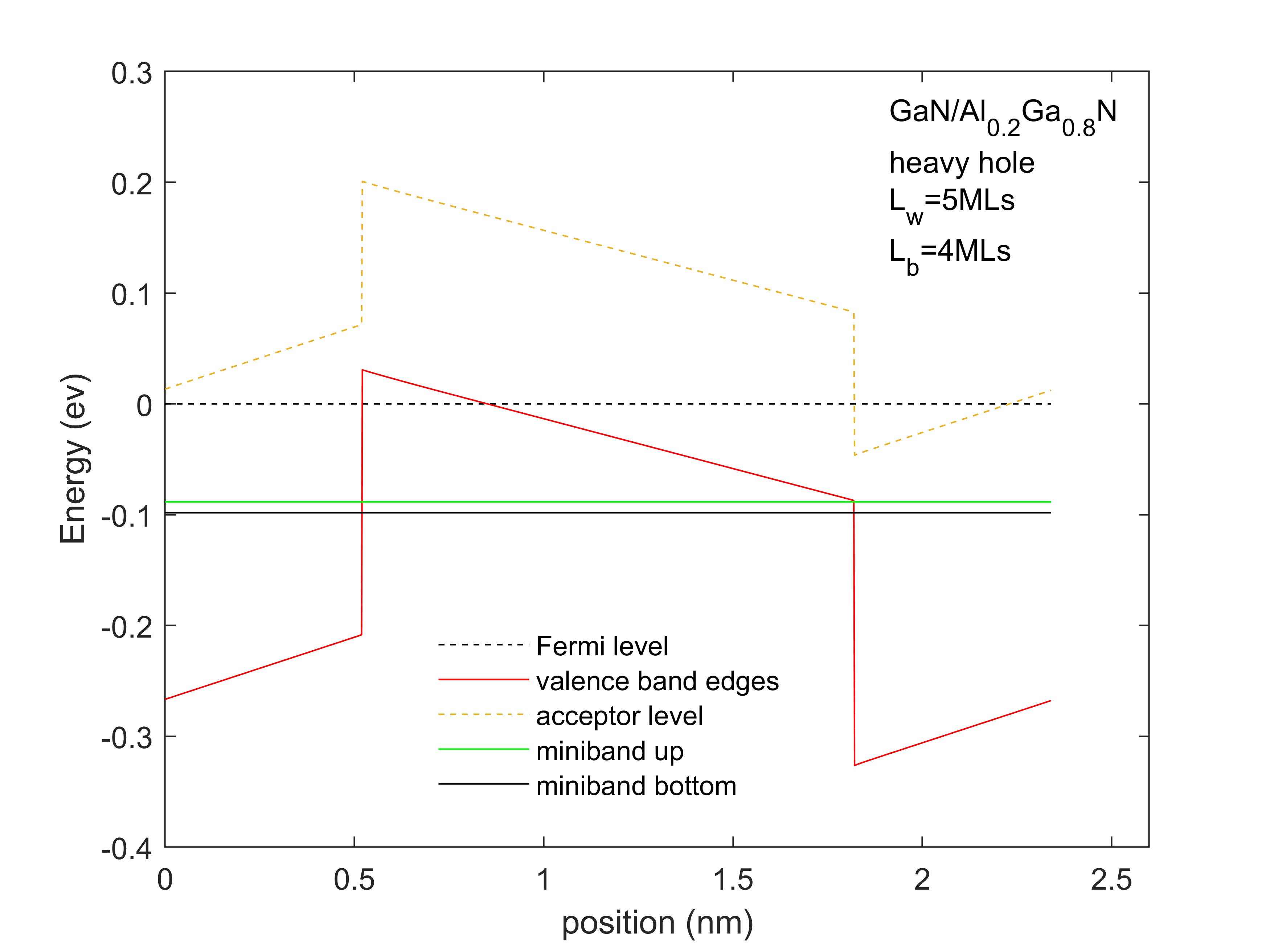}
    \caption{}
    \label{fig:first}
\end{subfigure}

\caption{The structure of GaN/Al$_{0.4}$Ga$_{0.6}$N 2MLs/2MLs (Fig.8a); GaN/Al$_{0.2}$Ga$_{0.8}$N 5MLs/4MLs (Fig.8b) without energy overlap. Doping concentration is $10^{20}cm^{-3}$.}
\label{fig:figures}
\end{figure}

In conclusion, we have proposed design criteria for a wurzite GaN/Al$_{x}$Ga$_{1-x}$N superlattice with a robust conducting channel and design options to minimize scattering from ionized impurities in the QBs. The SPS proposed have layer thicknesses of 4-6 MLs and low Al content so requires very good growth control. We show that achieving these targets is difficult and it may be the reason why GaN/Al$_{x}$Ga$_{1-x}$N superlattices have not demonstrated good vertical transport. It will be even more difficult to grow AlGaN based superlattices for use in UV LED and lasers as the effective mass will increase significantly in these SPS.
\newpage

\bibliography{sample}

\cite{Lambert:1989dg}.
\cite{lambert1989vertical}
\cite{deveaud1987bloch}
\cite{takahashi2007wide}
\cite{morkocc2013nitride}
\cite{liang2018progress}
\cite{duggan1998compound}
\cite{vurgaftman2001band}
\cite{duboz2014gan}
\cite{muhin2020vertical}
\cite{edmunds2013comparative}
\cite{hertkorn2008transport}
\cite{heikman2003high}
\cite{hess1979impurity}
\cite{arora1985phonon}
\cite{gantmakher2012carrier}
\cite{conwell1950theory}
\cite{friedman1985electron}
\cite{sztein2014polarization}
\cite{harrison2016quantum}
\cite{bank2001multigraph}
\cite{pampili2017doping}
\cite{park2000crystal}
\cite{park2000comparison}
\cite{zhao2017activation}
\cite{vurgaftman2007electron}
\cite{vurgaftman2003band}
\cite{sun2017aln}
\cite{o2006steady}
\cite{kozodoy1999enhanced}
\cite{ridley2013quantum}
\cite{dimitrijevprinciples}

\section*{Acknowledgements}
The authors would like to express their gratitude to the developers of nextnano software for providing a valuable tool in the simulation of our research results.

\section*{Author contributions statement}
Mengxun and Judy contributed to this work. Mengxun wrote the article as the first author. Both authors conceived and designed the simulation. Mengxun conducted the model, analyzed the data and performed the numerical simulations. Judy developed the theoretical framework as a supervisor. 

\section*{Data availability}
The datasets used and analyzed during the current study are available from the corresponding author upon reasonable request.

\end{document}